\documentclass[twocolumn,prc,showpacs,showkeys]{revtex4}
\usepackage{amsfonts}
\usepackage{amsmath}
\usepackage{amssymb}
\usepackage{graphicx}
\usepackage{rotating}

\providecommand{\U}[1]{\protect\rule{.1in}{.1in}}
\providecommand{\U}[1]{\protect\rule{.1in}{.1in}}
\providecommand{\U}[1]{\protect\rule{.1in}{.1in}}

\begin{document}

\title{Cooperative internal conversion process by proton exchange}
\author{P\'{e}ter K\'{a}lm\'{a}n\footnote{%
retired, e-mail: kalmanpeter3@gmail.com}}
\author{Tam\'{a}s Keszthelyi\footnote{%
retired, e-mail: khelyi@phy.bme.hu}}
\affiliation{Budapest University of Technology and Economics, Institute of Physics,
Budafoki \'{u}t 8. F., H-1521 Budapest, Hungary\ }

\begin{abstract}
A generalization of the recently discovered cooperative internal conversion
process is investigated theoretically. In the cooperative internal
conversion process by proton exchange investigated the coupling of
bound-free electron and proton transitions due to the dipole term of their
Coulomb interaction permits cooperation of two nuclei leading to proton
exchange and an electron emission. General expression of the cross section
of the process obtained in the one particle spherical nuclear shell model is
presented. As a numerical example the cooperative internal conversion
process by proton exchange in $Al$ is dealt with. As a further
generalization, cooperative internal conversion process by heavy charged
particle exchange and as an example of it the cooperative internal
conversion process by triton exchange is discussed. The process is also
connected to the field of nuclear waste disposal.
\end{abstract}

\volumenumber{number}
\issuenumber{number}
\eid{identifier}
\startpage{1}
\endpage{}
\maketitle

In a recent paper \cite{KK1} a new phenomenon, the cooperative internal
conversion process (CICP) is discussed which is a special type of the well
known internal conversion process \cite{Hamilton}. In CICP two nuclei
cooperate by neutron exchange creating final nuclei of energy lower than the
energy of the initial nuclei. The process is initiated by the coupling of
bound-free electron and neutron transitions due to the dipole term of their
Coulomb interaction in the initial atom leading to the creation of a virtual
free neutron which is captured through strong interaction by an other
nucleus. The process can be written as%
\begin{equation}
e+\text{ }_{Z_{1}}^{A_{1}}X+\text{ }_{Z_{2}}^{A_{2}}Y\rightarrow e^{\prime }+%
\text{ }_{Z_{1}}^{A_{1}-1}X+\text{ }_{Z_{2}}^{A_{2}+1}Y+\Delta ,
\label{exchange 0}
\end{equation}%
where $_{Z_{1}}^{A_{1}}X$, $_{Z_{2}}^{A_{2}}Y$ and $_{Z_{1}}^{A_{1}-1}X$, $%
_{Z_{2}}^{A_{2}+1}Y$ are the initial and final nuclei, respectively, and $e$
is an initial, bound electron of atom containing $_{Z_{1}}^{A_{1}}X$
nucleus, and $e^{\prime }$ is a free, final electron. $\Delta $ is the
energy of the reaction. $\Delta =\Delta _{-}+\Delta _{+},$ with $\Delta
_{-}=\Delta _{A_{1}}-\Delta _{A_{1}-1}$ and $\Delta _{+}=\Delta
_{A_{2}}-\Delta _{A_{2}+1}$. $\Delta _{A_{1}}$, $\Delta _{A_{1}-1}$, $\Delta
_{A_{2}}$, $\Delta _{A_{2}+1}$ are the energy excesses of neutral atoms of
mass numbers $A_{1}$, $A_{1}-1$, $A_{2}$, $A_{2}+1$, respectively \cite{Shir}%
. The process is mainly related to atomic state, i.e. it has accountable
cross section if the initial nuclei are located in free atoms therefore the
cross section was determined in the case of free atoms (e.g. noble gases) in
one particle nuclear and spherical shell models.

\begin{figure}[tbp]
\resizebox{8.0cm}{!}{\includegraphics{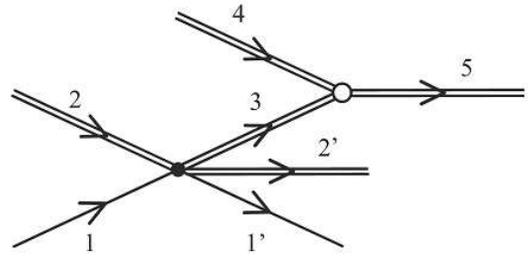}}
\caption{The graph of cooperative internal conversion process by heavy
charged particle (e.g. proton) exchange. Particle 1 (bound) and 1'(free) are
electrons, particle 2 is the nucleus which looses the heavy charged particle
(e.g. proton) and becomes particle 2'. Particle 3 is the intermediate heavy
charged particle (e.g. proton). Particle 4 is the nucleus which absorbs the
heavy charged particle (e.g. proton) and becomes particle 5. The filled dot
denotes (in case of proton the dipole term of) the Coulomb-interaction and
the open circle denotes nuclear (strong) interaction. }
\end{figure}

In this work the process

\begin{equation}
e_{1}+\text{ }_{Z_{1}}^{A_{1}}X+\text{ }_{Z_{2}}^{A_{2}}Y\rightarrow
e_{1}^{\prime }+\text{ }_{Z_{1}-1}^{A_{1}-1}V+\text{ }_{Z_{2}+1}^{A_{2}+1}W+%
\Delta ,  \label{exchange}
\end{equation}%
called cooperative internal conversion process by proton exchange (CICP-PE)
is discussed theoretically in more detail. In this process (see Fig. 1) a
bound proton of an atomic nucleus $\left( _{Z_{1}}^{A_{1}}X\text{, particle }%
2\right) $ is virtually excited into a free state (particle $3$) due to the
dipole term $V_{Cb}^{dip}$ of its Coulomb-interaction (in electric dipole
coupling the proton has effective charge $q_{p}=\left( 1-Z_{1}/A_{1}\right) e
$ \cite{Greiner}) with one of the bound atomic electrons $\left(
e_{1}\right) $ of the atom containing the $_{Z_{1}}^{A_{1}}X$ nucleus while
the electron becomes free $\left( e_{1}^{\prime }\right) $. The free,
virtual proton is captured by an other nucleus $_{Z_{2}}^{A_{2}}Y$ (particle 
$4$) due to its nuclear potential $V_{st}$ (created by strong interaction)
forming the nucleus $_{Z_{2}+1}^{A_{2}+1}W$ (particle $5$) in this way. The
sum of the rest energies of the initial nuclei is $E_{0i\text{ }}$ and the
sum of the the rest energies of the final nuclei is $E_{0f\text{ }}$. If $%
E_{0i\text{ }}-E_{0f\text{ }}=\Delta >0$, i.e. if $E_{0i\text{ }}>E_{0f\text{
}}$, then the process is energetically allowed and proton exchange is
possible. The nuclear energy difference $\Delta $, which is the reaction
energy, is shared between the kinetic energies of the final, free electron
and the two final nuclei [$_{Z_{1}-1}^{A_{1}-1}V$ (particle $2^{\prime }$)\
is the nucleus which has lost the proton].

The transition probability per unit time and the cross section $\sigma
_{bf}(A_{1},A_{2})$ of CICP-PE with bound-free $\left( bf\right) $ electron\
transitions can be determined with the aid of standard second order
perturbation calculation of quantum mechanics. The cross section has the
form $\sigma _{bf}(A_{1},A_{2})=\frac{c}{v}\sigma _{0bf}(A_{1},A_{2})$,
where $v$ is the relative velocity of the two atoms, $c$ is the velocity of
light (in vacuum). (The cross section of CICP-PE with bound-bound electron
transition is neglected since it has proved to be much smaller than $\sigma
_{bf}$.) The calculation of $\sigma _{0bf}$ is similar to the calculation of
CICP by neutron exchange \cite{KK1}. The differences are the appearance of
two Coulomb factors $F_{2^{\prime }3}$ and $F_{34}$ which multiply the cross
section and $\Delta _{n}$, $\frac{Z_{1}}{A_{1}}$ are changed to $\Delta _{p}$%
, $\left( 1-\frac{Z_{1}}{A_{1}}\right) $, respectively, in Eqs.(9)-(12) of 
\cite{KK1}. $\Delta _{p}=7.288969$ $MeV$ is the energy excess of the proton.
($F_{2^{\prime }3}$ and $F_{34}$ are determined in the Appendix.) Here we
repeat some essential features of the calculation. It is carried out in one
particle nuclear model. The motions of the centers of mass of the two atoms
are taken into account. Hydrogen like state of binding energy $E_{Bi}$ and
Coulomb-factor corrected plane wave are used as initial, bound and final,
free electron states. The dipole term of the Coulomb interaction reads as $%
V_{Cb}^{dip}=\left( 1-\frac{Z_{1}}{A_{1}}\right) e^{2}\frac{4\pi }{3}%
x_{1}x_{e}^{-2}\sum_{m=-1}^{m=1}Y_{1m}^{\ast }(\Omega _{e})Y_{1m}(\Omega
_{1})$, where $Z_{1}$ and $A_{1}$ are charge and mass numbers of the first
nucleus, $e$ is the elementary charge, $x_{1}$, $x_{e}$ and $\Omega _{1}$, $%
\Omega _{e}$ are magnitudes and solid angles of vectors $\mathbf{x}_{1}$, $%
\mathbf{x}_{e}$ which are the relative coordinates of the proton and the
electron in the first atom, respectively and $Y_{1m}$ denotes spherical
harmonics. (The order of magnitude of the cross section produced by the $L$%
-th pole coupling is $\left( R/r\right) ^{2L-2}$ times smaller than the
cross section produced by the dipole coupling where $R$ and $r$ are the
nuclear and atomic radii. Therefore the leading term to the cross section is
produced by the dipole coupling.) The motion of the intermediate proton and
the two final nuclei are also described by plane waves. The rest masses of
the two initial nuclei of mass numbers $A_{1}$ and $A_{2}\ $are $%
m_{1}=A_{1}m_{0}$ (particle $2$) and $m_{2}=A_{2}m_{0}$ (particle $4$) where 
$m_{0}c^{2}=931.494$ $MeV$ is the atomic energy unit. For the nuclear
potential a rectangular potential well is assumed, i.e. $V_{st}=-V_{0}$ $%
(x_{2}\leq R_{A_{2}})$ and $V_{st}=0$ $(x_{2}>R_{A_{2}})$ where $x_{2}$ is
the magnitude of vector $\mathbf{x}_{2}$, which is the relative coordinate
of the neutron in the second nucleus and $R_{A_{2}}$ is its radius. Direct
proton capture may be assumed at the surface of the second nucleus (of $%
A_{2} $). The effective volume in which strong interaction induces proton
capture can be considered as a shell of a sphere of radius $R_{A_{2}}$ and
of thickness $L$, where $L$ is the mean free path of the ingoing proton in
the nucleus \cite{Blatt}.

Introducing the wave vectors $\mathbf{k}_{e}$ and $\mathbf{k}_{1}$, $\mathbf{%
k}_{2}$ of the free electron and particles $_{Z_{1}-1}^{A_{1}-1}V$ (particle 
$2^{\prime }$) and $_{Z_{2}+1}^{A_{2}+1}W$ (particle $5$), respectively, the
analysis of $\sigma _{bf}$ shows that, similarly to the CICP by neutron
exchange \cite{KK1}, those processes give essential contribution to the
cross section in which $k_{e}\ll $ $k_{1}$ and $k_{e}\ll $ $k_{2}$ where $%
k_{e}$, $k_{1}$ and $k_{2}$ are the magnitudes of the wave vectors of $%
\mathbf{k}_{e}$, $\mathbf{k}_{1}$ and $\mathbf{k}_{2}$. (In this case as a
consequence of momentum conservation $\mathbf{k}_{1}=-\mathbf{k}_{2}$,
furthermore the intermediate proton has wave vector $-\mathbf{k}_{2}$.)

The initial and final nuclear states have the form: $\phi _{i}\left( \mathbf{%
x}_{1}\right) =\varphi _{i}\left( x_{1}\right) Y_{l_{i}m_{i}}(\Omega
_{1})/x_{1}$ and $\phi _{f}\left( \mathbf{x}_{2}\right) =\varphi _{f}\left(
x_{2}\right) Y_{l_{f}m_{f}}(\Omega _{2})/x_{2}$ with $\varphi _{i}\left(
x_{1}\right) /x_{1}$ and $\varphi _{f}\left( x_{2}\right) /x_{2}$ denoting
the radial parts of the one particle shell-model solutions of quantum
numbers $l_{i}$, $m_{i}$ and $l_{f}$, $m_{f}$. For $\varphi _{i}\left(
x_{1}\right) $ and $\varphi _{f}\left( x_{2}\right) $ the corresponding part 
$R_{0\Lambda }=b_{k}^{-1/2}\Gamma (\Lambda +3/2)^{-1/2}2^{1/2}\rho
_{k}^{\Lambda +1}\exp (-\rho _{k}^{2}/2)$ of the $0\Lambda $ one particle
spherical shell model states \cite{Pal} is applied. Here $\rho
_{k}=x_{k}/b_{k}$, $b_{k}=\left( \frac{\hbar }{m_{0}\omega _{sh,k}}\right)
^{1/2}$ and $\hbar \omega _{sh,k}=41A_{k}^{-1/3}$ (in $MeV$ units, \cite%
{Greiner}) with $k=1,2$ corresponding to $A_{1}$ and $A_{2}$, and $\Gamma
(x) $ is the gamma function. The case of spherical shell model states of $%
0l_{i}$ initial nuclear state and of $0l_{f}$ final nuclear state is
investigated.

The initial electronic state is a$\ 1s$ state of the form $R_{i}\left(
x_{e}\right) =2a^{-3/2}\exp (-x_{e}/a)$ with $a=a_{0}/Z_{eff}$, where $a_{0}$
is the Bohr-radius, $Z_{eff}=\sqrt{E_{B}/Ry}$ and $Ry$ is the Rydberg
energy. In the Coulomb-corrected plane wave applied for the final free
electron the $F_{Cb}\left( k_{e}\right) =2\pi /\left( k_{e}a\right) $
approximation is used, where $F_{Cb}\left( k_{e}\right) $ is the Coulomb
factor of the electron. Keeping the leading term of $J_{e}^{1}(k_{e})$ in 
\cite{KK1} and in the case of $l_{i}=even$ [$l_{i}=2$; $Al(5/2^{+},0d)$] to
be investigated one obtains 
\begin{eqnarray}
\sigma _{0bf,sh} &=&\frac{2^{10}\pi ^{3}}{3}\left( 1-\frac{Z_{1}}{A_{1}}%
\right) ^{2}\frac{V_{0}^{2}}{\left( \hbar c\right) ^{2}}\frac{b_{1}^{5}L^{2}%
}{\lambdabar _{e}a_{0}^{2}}\frac{m_{0}}{m_{e}}a_{12}  \label{sigma03} \\
&&\times \left( 2l_{f}+1\right) \frac{\rho _{f}^{2l_{f}+3}e^{-\rho _{f}^{2}}%
}{\Gamma \left( l_{f}+\frac{1}{2}\right) }\sum_{\lambda =l_{i}\pm 1}\frac{%
N_{1\lambda }\left( k_{0}b_{1}\right) ^{2\lambda }}{\Gamma \left( \lambda +%
\frac{3}{2}\right) }S_{\lambda }.  \notag
\end{eqnarray}%
Here $\lambdabar _{e}=\hbar /(m_{e}c)$, $m_{e}$ is the rest mass of the
electron, $a_{12}=\left( A_{1}-1\right) \left( A_{2}+1\right) /\left(
A_{1}+A_{2}\right) $, $\rho _{f}=R_{A_{2}}/b_{2}$, $k_{0}=\sqrt{2m_{0}\Delta
_{Bi}a_{12}}/\hbar $, and 
\begin{equation}
N_{1\lambda }=(2\lambda +1)\left( 
\begin{array}{ccc}
l_{i} & 1 & \lambda \\ 
0 & 0 & 0%
\end{array}%
\right) ^{2}.  \label{N01la}
\end{equation}%
The parenthesized expression is Wigner 3j symbol. (The suffix $sh$ denotes
that the quantity is calculated in the one particle spherical shell model.) 
\begin{equation}
S_{\lambda }=\int_{0}^{1}f\left( x\right) g_{\lambda }\left( x\right)
h_{1}\left( x\right) h_{2}\left( x\right) dx,\text{ }  \label{slam}
\end{equation}%
\begin{equation}
f\left( x\right) =\frac{\left( 1-x^{2}\right) x^{2\lambda
+1}e^{-(k_{0}b_{1})^{2}x^{2}}J_{l_{f}+\frac{1}{2}}^{2}(xk_{0}R_{A_{2}})}{%
\left[ 1+\frac{\Delta _{Bi}}{E_{B}}\left( 1-x^{2}\right) \right] ^{2}\left[ 
\frac{A_{1}a_{12}}{A_{1}-1}x^{2}+1+\xi \right] ^{2}},  \label{fkszi}
\end{equation}%
$x=k_{2}/k_{0}$, $\xi =\left( \Delta _{p}-\Delta _{-}+E_{Bi}\right) /\Delta
_{Bi}$ and $\Delta _{Bi}=\Delta -E_{Bi}$. $J_{l_{f}+\frac{1}{2}}$ is a
Bessel-function of the first kind. In Eq.$\left( \ref{slam}\right) $ $%
g_{\lambda }\left( x\right) =1$ if $\lambda =l_{i}+1$ and 
\begin{equation}
g_{\lambda }\left( x\right) =\left( 2l_{i}+1\right) ^{2}-2\left(
2l_{i}+1\right) (k_{0}b_{1}x)^{2}+(k_{0}b_{1}x)^{4}  \label{glak2}
\end{equation}%
if $\lambda =l_{i}-1$. $h_{j}\left( x\right) =d_{j}(x)/\left[ \exp
(d_{j}(x))-1\right] $, $j=1,2$ with%
\begin{equation}
d_{1}(x)=2\pi (Z_{1}-1)\alpha _{f}\frac{1}{x}\sqrt{\frac{\left(
A_{1}+A_{2}\right) m_{0}c^{2}}{2A_{1}\left( A_{2}+1\right) \Delta _{Bi}}}
\label{d1}
\end{equation}%
and 
\begin{equation}
d_{2}(x)=2\pi Z_{2}\alpha _{f}\frac{1}{x}\sqrt{\frac{\left(
A_{1}+A_{2}\right) m_{0}c^{2}}{2\left( A_{1}-1\right) \left( A_{2}+1\right)
\Delta _{Bi}}}.  \label{d2}
\end{equation}%
$\alpha _{f}$ denotes the fine structure constant. In the numerical
calculation $V_{0}=50$ $MeV$ is used \cite{Greiner}.

The differential cross section $d\sigma _{0bf,sh}/dE_{2}$ of the process can
be determined with the aid of 
\begin{equation}
P(x)=\sum_{\lambda =l_{i}\pm 1}\frac{N_{1\lambda }\left( k_{0}b_{1}\right)
^{2\lambda }}{\Gamma \left( \lambda +\frac{3}{2}\right) }\frac{f\left(
x\right) g_{\lambda }\left( x\right) h_{1}\left( x\right) h_{2}\left(
x\right) }{x}  \label{Gy}
\end{equation}%
as $d\sigma _{0bf,sh}/dE_{2}=K_{bf}\left[ P(x)\right] _{x=\sqrt{z}}/\left(
2E_{20}\right) $ where $z=E_{2}/E_{20}$ with $E_{20}=\left( A_{1}-1\right)
\Delta _{Bi}/\left( A_{1}+A_{2}\right) $, which is the possible maximum of
the kinetic energy $E_{2}$ of particle $_{Z_{2}+1}^{A_{2}+1}W$ (particle $5$%
) created in the process, $K_{bf}$ stands for the whole factor which
multiplies the sum in $\left( \ref{sigma03}\right) $. $d\sigma
_{0bf,sh}/dE_{2}$ has accountable values near below $z=1$, i.e. if $%
E_{2}\sim E_{20}$.

The differential cross section $d\sigma _{0bf,sh}/dE_{e}=K_{bf}\left[ P(x)%
\right] _{x=\sqrt{1-z}}/\left( 2\Delta _{Bi}\right) $ can also be determined
with the aid of $P(x)$ where $z=E_{e}/\Delta _{Bi}$, $E_{e}$ is the kinetic
energy of the electron and $K_{bf}$ is defined above. $d\sigma
_{0bf,sh}/dE_{e}$ has accountable values near above $z=0$, i.e. if $%
E_{e}\sim 0$.

It is a special case of $\left( \ref{exchange}\right) $ if the two initial
nuclei are identical. In this case the CICP-PE reads as%
\begin{equation}
e_{1}+\text{ }_{Z_{1}}^{A_{1}}X+\text{ }_{Z_{1}}^{A_{1}}X\rightarrow
e_{1}^{\prime }+\text{ }_{Z_{1}-1}^{A_{1}-1}V+\text{ }_{Z_{1}+1}^{A_{1}+1}W+%
\Delta .  \label{exchange2}
\end{equation}%
For example of such a case the reaction $e+$ $_{13}^{27}Al+$ $%
_{13}^{27}Al\rightarrow e^{\prime }+$ $_{12}^{26}Mg+$ $_{14}^{28}Si+\Delta $
is considered when the reaction starts from the $K$ shell. The initial and
final nuclear states are supposed to be $0d$ spherical shell model states of 
$l_{i}=l_{f}=2$, $\Delta =3.31362$ $MeV$. The electron binding energy in the 
$K$ shell is $E_{Bi}=1.5596$ $keV$ and $\Delta _{-}=-0.98235$ $MeV$. In this
case $2E_{20}=3.1894$ $MeV$ and $K_{bf}/\left( 2E_{20}\right) =2.41\times
10^{-35}$ $cm^{2}MeV^{-1})$. $\sigma _{0bf,sh}(K)=2.41\times 10^{-46}$ $%
cm^{-2}$ is obtained in the case of bound-free CICP from the $K$ shell of $%
Al $. If one compares this result with $\sigma _{0bf,sh}(K)=$ $8.25\times
10^{-45}$ $cm^{-2}$obtained in case of CICP by neutron exchange in $Ne$ one
can recognize that the ratio of the two cross sections is only $0.030$. At
first sight it seems to be larger when expected since two Coulomb factors
appeare in the cross section. But as it was said earlier the intermediate
proton has wave vector $-\mathbf{k}_{2}$ and thus its energy $E_{3}=\hbar
^{2}\mathbf{k}_{2}^{2}/\left( 2m_{0}\right) $ with $\hbar k_{2}=\hbar
k_{0}x=x\sqrt{2m_{0}\Delta _{Bi}a_{12}}$. It gives $E_{3}=x^{2}\Delta
_{Bi}A_{1}/2$ since $a_{12}=A_{1}/2$ if $A_{1}=A_{2}$ and near below $x=1$
the value of $E_{3}$ is large enough to result moderately small Coulomb
factors.

For a gas of atomic $Al$ and of number density $n$ the transition
probability per unit time $\lambda _{1}=cn\sum_{A_{2}}r_{A_{2}}\sigma
_{0bf,sh}=cn\sigma _{0bf,sh}$ \cite{KK1} since the relative natural
abundance $r_{A_{2}}$ of the initial $_{13}^{27}Al$ isotope equals unity. $%
\lambda _{1}$\ is estimated as $\lambda _{1}>\lambda _{1}(K)$, which is the
transition probability per unit time of the bound-free CICP-PE from the $K$
shell of $Al$ ($\lambda _{1}(K)=cn\sigma _{0bf,sh}(K)$), resulting $\lambda
_{1}>1.92\times 10^{-16}$ $s^{-1}$ and $r_{tot}>5.09\times 10^{3}$ $%
cm^{-3}s^{-1}$ for a gas of normal state ($n=2.652\times 10^{19}cm^{-3}$, $%
T=273.15$ $K$, $p=100$ $kPa$ and $r_{tot}=n\lambda _{1}$, which is the total
rate per unit volume of the sample, in this case since $r_{A_{1}}=1$ \cite%
{KK1}).

\begin{table}[tbp]
\tabskip=8pt 
\centerline {\vbox{\halign{\strut $#$\hfil&\hfil$#$\hfil&\hfil$#$
\hfil&\hfil$#$\hfil&\hfil$#$\hfil&\hfil$#$\cr
\noalign{\hrule\vskip2pt\hrule\vskip2pt}
Isotope&Products&\Delta_{-} (MeV)&\Delta_{+} (MeV)&\Delta (MeV) \cr
\noalign{\vskip2pt\hrule\vskip2pt}
^{19}F & ^{18}O,  ^{20}Ne & -0.705 & 5.555 & 4.850 \cr
^{23}Na & ^{22}Ne,  ^{24}Mg & -1.505 &4.404 & 2.899 \cr
^{27}Al & ^{26}Mg,  ^{28}Si & -0.982 & 4.296 & 3.314 \cr
^{31}P & ^{30}Si,  ^{32}S & -0.008 & 1.575 & 1.567 \cr
^{45}Sc & ^{44}Ca,  ^{46}Ti  & 0.400 & 3.056 & 3.456 \cr
^{55}Mn & ^{54}Cr,  ^{56}Fe  & -0.778 & 2.895 & 2.117 \cr
^{59}Co & ^{58}Fe,  ^{60}Ni  & -0.075 & 2.245 & 2.170 \cr
^{103}Rh & ^{102}Ru,  ^{104}Pd & 1.076 & 1.369 & 2.445 \cr
^{127}I & ^{126}Te,  ^{128}Xe & 1.083 & 0.873 & 1.956 \cr
^{133}Cs & ^{132}Xe,  ^{134}Ba & 1.204 & 0.879 & 2.083 \cr
\noalign{\vskip2pt\hrule\vskip2pt\hrule}}}}
\caption{Data for cooperative internal conversion process by proton
exchange. (Data to reaction $\left( \protect\ref{exchange2}\right) $.) In
the first column the initial stable isotope (of unity relative natural
abundance) and in the second column the reaction products can be found. For
the definition of $\Delta _{-}$, $\Delta _{+}$ and $\Delta $ see the text.}
\label{Table1}
\end{table}
In Table I. the $\Delta _{-}$, $\Delta _{+}$ and $\Delta $ data of some
cooperative internal conversion processes by proton exchange (data to
reaction $\left( \ref{exchange2}\right) $ can be found. In the first column
the initial stable isotope of relative natural abundance unity and in the
second column the reaction products are listed.

There are other possibilities to realize CICP, when a charged heavy particle
(such as $d$, $t$, $_{2}^{3}He$ and $_{2}^{4}He$) is exchanged instead of
proton exchange. The process is called cooperative internal conversion
process by heavy charged particle exchange (CICP-HCPE) and it can be
visualized with the aid of Fig.1 too. Denoting the intermediate particle
(particle $3$ in Fig. 1) by $_{z_{3}}^{A_{3}}w$, which is exchanged, the
cooperative internal conversion process by heavy charged particle exchange
reads as%
\begin{equation}
e+\text{ }_{Z_{1}}^{A_{1}}X+\text{ }_{Z_{2}}^{A_{2}}Y\rightarrow e^{\prime }+%
\text{ }_{Z_{1}-z_{3}}^{A_{1}-A_{3}}V+\text{ }_{Z_{2}+z_{3}}^{A_{2}+A_{3}}W+%
\Delta .  \label{hpexchange}
\end{equation}%
Here $e$ and $e^{\prime }$ denote bound and free electron and $\Delta $ is
the energy of the reaction, i.e. the difference between the rest energies of
initial $\left( _{Z_{1}}^{A_{1}}X+_{Z_{2}}^{A_{2}}Y\right) $ and final $%
\left( _{Z_{1}-z_{3}}^{A_{1}-A_{3}}V+\text{ }_{Z_{2}+z_{3}}^{A_{2}+A_{3}}W%
\right) $ states. $\Delta =\Delta _{-}+\Delta _{+},$ with $\Delta
_{-}=\Delta _{Z_{1}}^{A_{1}}-\Delta _{Z_{1}-z_{3}}^{A_{1}-A_{3}}$ and $%
\Delta _{+}=\Delta _{Z_{2}}^{A_{2}}-\Delta _{Z_{2}+z_{3}}^{A_{2}+A_{3}}$. $%
\Delta _{Z_{1}}^{A_{1}}$, $\Delta _{Z_{1}-z_{3}}^{A_{1}-A_{3}}$ , $\Delta
_{Z_{2}}^{A_{2}}$, $\Delta _{Z_{2}+z_{3}}^{A_{2}+A_{3}}$ are the energy
excesses of neutral atoms of mass number-charge number pairs $A_{1}$, $Z_{1}$%
; $A_{1}-A_{3}$, $Z_{1}-z_{3}$; $A_{2}$, $Z_{2}$; $A_{2}+A_{3}$, $%
Z_{2}+z_{3} $, respectively \cite{Shir}.

In $\left( \ref{hpexchange}\right) $ the initial bound electron (particle $1$%
) Coulomb interacts with the nucleus $_{Z_{1}}^{A_{1}}X$ (particle $2$). A
free electron (particle $1^{\prime }$), the intermediate particle $%
_{z_{3}}^{A_{3}}w$ (particle $3$) and the nucleus $%
_{Z_{1}-z_{3}}^{A_{1}-A_{3}}V$ (particle $2^{\prime }$) are created due to
this interaction. The intermediate particle $_{z_{3}}^{A_{3}}w$ (particle $3$%
) is captured due to the strong interaction by the nucleus $%
_{Z_{2}}^{A_{2}}Y $ (particle $4$) forming the nucleus $%
_{Z_{2}+z_{3}}^{A_{2}+A_{3}}W$ (particle $5$) in this manner. So in $\left( %
\ref{hpexchange}\right) $ the nucleus $_{Z_{1}}^{A_{1}}X$ (particle $2$)
looses a particle $_{z_{3}}^{A_{3}}w$ which is taken up by the nucleus $%
_{Z_{2}}^{A_{2}}Y$ (particle $4$). The process is energetically forbidden if 
$\Delta <0$.

As an example we deal with the 
\begin{equation}
e+\text{ }_{Z_{1}}^{A_{1}}X+\text{ }_{Z_{2}}^{A_{2}}Y\rightarrow e^{\prime }+%
\text{ }_{Z_{1}-1}^{A_{1}-3}V+\text{ }_{Z_{2}+1}^{A_{2}+3}W+\Delta
\label{hpexchange2}
\end{equation}%
reaction in which a triton is exchanged. It is called CICP by triton
exchange (CICP-TE). Special type of reaction $\left( \ref{hpexchange2}%
\right) $ is 
\begin{equation}
e+\text{ }_{Z_{1}}^{A_{1}}X+\text{ }_{Z_{1}}^{A_{1}}X\rightarrow e^{\prime }+%
\text{ }_{Z_{1}-1}^{A_{1}-3}V+\text{ }_{Z_{2}+1}^{A_{2}+3}W+\Delta .
\label{hpexchange3}
\end{equation}%
In Table II. the $\Delta _{-}$, $\Delta _{-}$ and $\Delta $ data of some
cooperative internal conversion processes by triton exchange (data to
reaction $\left( \ref{hpexchange3}\right) $ can be found. 
\begin{table}[tbp]
\tabskip=8pt 
\centerline {\vbox{\halign{\strut $#$\hfil&\hfil$#$\hfil&\hfil$#$
\hfil&\hfil$#$\hfil&\hfil$#$\hfil&\hfil$#$\cr
\noalign{\hrule\vskip2pt\hrule\vskip2pt}
Isotope&Products&\Delta_{-} (MeV)&\Delta_{+} (MeV)&\Delta (MeV) \cr
\noalign{\vskip2pt\hrule\vskip2pt}
^{19}F & ^{16}O, ^{22}Ne & 3.250 & 6.537 & 9.787 \cr
^{23}Na & ^{20}Ne, ^{26}Mg & -2.488 & 6.685 & 4.197 \cr
^{27}Al & ^{24}Mg, ^{30}Si & -3.263 & 7.236 & 3.973 \cr
^{31}P & ^{28}Si, ^{34}S & -2.948 & 5.491 & 2.543 \cr
^{45}Sc & ^{42}Ca, ^{48}Ti  & -2.522 & 7.418 & 4.896 \cr
^{55}Mn & ^{52}Cr, ^{58}Fe  & -2.294 & 4.443 & 2.149 \cr
^{59}Co & ^{56}Fe, ^{62}Ni  & -1.623 & 4.519 & 2.896 \cr
^{103}Rh & ^{100}Ru, ^{106}Pd & 1.197 & 1.882 & 3.079 \cr
^{127}I & ^{124}Te, ^{130}Xe & 1.536 & 0.893 & 2.429 \cr
^{133}Cs & ^{130}Xe, ^{136}Ba & 1.806 & 0.816 & 2.622 \cr
\noalign{\vskip2pt\hrule\vskip2pt\hrule}}}}
\caption{Data for cooperative internal conversion process by triton
exchange. (Data to reaction $\left( \protect\ref{hpexchange3}\right) $.) In
the first column the initial stable isotope (of unity relative natural
abundance) and in the second column the reaction products can be found. For
the definition of $\Delta _{-}$, $\Delta _{+}$ and $\Delta $ see the text.}
\label{Table2}
\end{table}

\begin{table}[tbp]
\tabskip=8pt 
\centerline {\vbox{\halign{\strut $#$\hfil&\hfil$#$\hfil&\hfil$#$
\hfil&\hfil$#$\hfil&\hfil$#$\hfil&\hfil$#$\cr
\noalign{\hrule\vskip2pt\hrule\vskip2pt}
Isotope&\tau (y)&Products&\Delta_{-} (MeV)&\Delta_{+} (MeV) \cr
\noalign{\vskip2pt\hrule\vskip2pt}
^{99}Tc & 2.11\times 10^{5} & ^{98}Mo, ^{100}Ru & 0.789 & 1.896 \cr
^{129}I & 1.57\times 10^{7} & ^{128}Te, ^{130}Xe  & 0.491 & 1.378 \cr
^{135}Cs & 2.3\times 10^{6} & ^{134}Xe, ^{136}Ba & 0.538 & 1.305 \cr
^{137}Cs & 30.07 & ^{136}Xe, ^{138}Ba & -0.126 & 1.716 \cr
^{155}Eu & 4.7611 & ^{154}Sm, ^{156}Gd & 0.637 & 0.717 \cr
\noalign{\vskip2pt\hrule\vskip2pt\hrule}}}}
\caption{Data for cooperative internal conversion process by proton exchange
of long lived nuclear fission products. (Data to reaction $\left( \protect
\ref{exchange2}\right) $.) Products are the two stable final izotopes, $%
\protect\tau $ is the half-life of the fission product in $y$ units. For the
definition of $\Delta _{-}$ and $\Delta _{+}$ see the text. }
\label{Table3}
\end{table}
In Table III. some long lived fission products are listed which can take
part in CICP-PE. The values of $\Delta _{-}$ and $\Delta _{-}$ indicate that
each pair of the listed isotopes can produce CICP-PE since $\Delta
_{-}+\Delta _{-}$ $=\Delta >0$ in every case. Consequently similarly to
those long lived fission products which can decay through CICP by neutron
exchange \cite{KK1}, it seems to stand also a practical chance to accelerate
the decay of the listed isotopes if they were collected in appropriately
high concentration and density in atomic state.

\subsection{Appendix - Coulomb factors $F_{2^{\prime }3}$ and $F_{34}$}

Since particles $2^{\prime }$, $3$ and $4$ all have positive charge,
furthermore they are all heavy, the two essential Coulomb factors, which
appear in the cross section, are $F_{2^{\prime }3}$ and $F_{34}$. Since
Coulomb factors $F_{2^{\prime }3}$ and $F_{34}$ determine the order of
magnitude of the cross section of the process (as it is proportional to $%
F_{2^{\prime }3}F_{34}$) we treat them in more detail in the case of CICP-PE
in the following.

We adopt the approach standard in nuclear physics when describing the cross
section of nuclear reactions of heavy, charged particles $j$ and $k$ of like
positive charge of charge numbers $z_{j}$ and $z_{k}$ and of relative
kinetic energy $E$. The cross section of such a process can be\ derived
applying the Coulomb solution $\varphi (\mathbf{r})$, 
\begin{equation}
\varphi (\mathbf{r})=e^{i\mathbf{k}\cdot \mathbf{r}}f(\mathbf{k,r})/\sqrt{V},
\label{Cb1}
\end{equation}%
which is the wave function of a free particle of charge number $z_{j}$ in a
repulsive Coulomb field of charge number $z_{k}$ \cite{Alder}, in the
description of relative motion of projectile and target. In $\left( \ref{Cb1}%
\right) $ $V$ denotes the volume of normalization, $\mathbf{r}$ is the
relative coordinate of the two particles, $\mathbf{k}$ is the wave number
vector in their relative motion and 
\begin{equation}
f(\mathbf{k},\mathbf{r})=e^{-\pi \eta _{jk}/2}\Gamma (1+i\eta
_{jk})_{1}F_{1}(-i\eta _{jk},1;i[kr-\mathbf{k}\cdot \mathbf{r}]),
\label{Hyperg}
\end{equation}%
where $_{1}F_{1}$ is the confluent hypergeometric function and $\Gamma $ is
the Gamma function. Since $\varphi (\mathbf{r})\sim e^{-\pi \eta
_{jk}/2}\Gamma (1+i\eta _{jk})$, the cross section of the process is
proportional to 
\begin{equation}
\left\vert e^{-\pi \eta _{jk}/2}\Gamma (1+i\eta _{jk})\right\vert ^{2}=\frac{%
2\pi \eta _{jk}\left( E\right) }{\exp \left[ 2\pi \eta _{jk}\left( E\right) %
\right] -1}=F_{jk}(E),  \label{Fjk}
\end{equation}%
which is the so-called Coulomb factor. Here 
\begin{equation}
\eta _{jk}\left( E\right) =z_{j}z_{k}\alpha _{f}\sqrt{a_{jk}\frac{m_{0}c^{2}%
}{2E}}  \label{etajk}
\end{equation}%
is the Sommerfeld parameter in the case of colliding particles of mass
numbers $A_{j}$, $A_{k}$ and rest masses $m_{j}=A_{j}m_{0}$, $%
m_{k}=A_{k}m_{0}$. $m_{0}c^{2}=931.494$ $MeV$ is the atomic energy unit, $%
\alpha _{f}$ is the fine structure constant and $E$ \ is taken in the center
of mass $\left( CM\right) $ coordinate system.%
\begin{equation}
a_{jk}=\frac{A_{j}A_{k}}{A_{j}+A_{k}}  \label{ajk}
\end{equation}%
is the reduced mass number of particles $j$ and $k$ of mass numbers $A_{j}$
and $A_{k}$.

If initial particles have negligible initial momentum then in the final
state $\mathbf{k}_{1}=-\mathbf{k}_{2}$ ($\mathbf{k}_{particle,2^{\prime }}=-%
\mathbf{k}_{particle,5}$) because of momentum conservation. (It was obtained 
\cite{KK1} that the process has accountable cross section if the momentum of
the final electron can be neglected.) In this case the momentum \ and energy
of the virtual particle $3$ (e.g. proton) are $\mathbf{k}_{particle,3}=-%
\mathbf{k}_{particle,2^{\prime }}=\mathbf{k}_{particle,5}\equiv $ $\mathbf{k}%
_{2}$ and $E_{3}=\hbar ^{2}\mathbf{k}_{2}^{2}/\left( 2m_{3}\right) $, where $%
\hbar $ is the reduced Planck-constant. Calculating the Coulomb factor $%
F_{2^{\prime }3}$ [see $\left( \ref{Fjk}\right) $]\ between particles $%
2^{\prime }$ and $3$ the energy $E_{3}$ is given in their $CM$ coordinate
system (since $\mathbf{k}_{particle,3}=-\mathbf{k}_{particle,2^{\prime }}$)
thus $E_{3}$ can be substituted directly in $\left( \ref{etajk}\right) $
producing%
\begin{equation}
\eta _{2^{\prime }3}=\left( Z_{1}-1\right) \alpha _{f}\frac{1}{x}\sqrt{\frac{%
A_{1}+A_{2}}{A_{1}\left( A_{2}+1\right) }\frac{m_{0}c^{2}}{2\Delta _{Bi}}}.
\label{eta2'3}
\end{equation}%
in case of proton exchange. Here the $k_{2}=k_{0}x$ substitution is also
used. Calculating the Coulomb factor $F_{34}$, the energy $E_{3}$ of
particle $3$ is now given in the laboratory frame of reference since
particle $4$ is at rest. In the $CM$ system of particles $3$ and $4$ the
energy $E_{3}(CM)$ is 
\begin{equation}
E_{3}(CM)=\frac{A_{particle,4}}{\left( A_{particle,3}+A_{particle,4}\right) }%
E_{3}.  \label{E3CM}
\end{equation}%
Substituting it into $\left( \ref{etajk}\right) $ 
\begin{equation}
\eta _{34}=Z_{2}\alpha _{f}\frac{1}{x}\sqrt{\frac{\left( A_{1}+A_{2}\right)
m_{0}c^{2}}{2\left( A_{1}-1\right) \left( A_{2}+1\right) \Delta _{Bi}}}
\label{eta34}
\end{equation}%
in case of proton exchange.

\end{document}